\documentclass[prb,twocolumn,showpacs,preprintnumbers,amsmath,amssymb]{revtex4}
\usepackage{amssymb}
\usepackage{bm}
\usepackage{amsmath}
\usepackage[dvips]{graphicx}
\usepackage{color}

\begin{document}

\title{
Magnetic properties of lightly doped antiferromagnetic 
YBa$_{2}$Cu$_{3}$O$_{y}$.
}

\author{O. P. Sushkov}
\affiliation{School of Physics, University of New South Wales, Sydney 2052, Australia}

\begin{abstract}
The present work addresses YBa$_{2}$Cu$_{3}$O$_{y}$ at doping below
$x=6\%$ where the compound is a collinear antiferromagnet. 
In this region YBa$_{2}$Cu$_{3}$O$_{y}$ is a normal 
conductor with a finite  resistivity at zero temperature.
The value of the staggered magnetization at zero temperature is
$\approx 0.6\mu_B$, the maximum value allowed by spin quantum fluctuations.
The staggered magnetization is almost independent of doping.
On the other hand, the Neel temperature decays very quickly from 
$T_N=420K$ at $x=0$ to practically zero at  $x \approx 6\%$.
The present paper explains these remarkable properties and
demonstrates that the properties
 result from the physics of a lightly doped 
Mott insulator with small hole pockets. 
Nuclear quadrupole resonance data are also discussed.
The data shed light on
mechanisms of stability of the antiferromagnetic order at $x < 6\%$.
\end{abstract}

\date{\today}
\maketitle

\section{Introduction}
It is well known that cuprates are layered
compounds consisting of CuO$_2$ planes
and there are no doubts that the generic physics of 
cuprates are related to the CuO$_2$ plane.
In spite of  the same generic physics
specific properties of cuprates can be very different depending on
crystal structure, ways of doping etc. 
The goal of the present work it to shed light on the generic physics
via understanding of specific properties  of
lightly doped antiferromagnetic YBa$_{2}$Cu$_{3}$O$_{y}$ (YBCO). 

Cuprates are essentially doped Mott insulators. It is well established that
a Mott insulator possesses a long range antiferromagnetic (AF) order,
therefore, one of the generic problems is how the AF order evolves with doping.
Another generic problem is the shape of the Fermi surface. Are there
small hole pockets as one  expects for a very lightly doped Mott
insulator, how the surface evolves with doping? 

Cuprates are intrinsically disordered materials because of
 mechanisms of doping.  Disorder complicates a theoretical analysis 
of experimental data  usually  masking the generic
physics. YBCO is probably the least disordered cuprate
in the low doping regime. In this paper I denote the hole concentration
per unit cell of the CuO$_2$ layer  by $x$, this is the ``doping''. YBCO is not
superconducting below $x\approx 0.06$ where it behaves as a normal
conductor with delocalized holes.
 The zero temperature resistivity remains finite,~\cite{WO01}
apart of a very weak logarithmic temperature dependence~\cite{sun,doiron}
expected for a weak disorder. 
The heat conductivity also indicates delocalization of 
holes.~\cite{sutherland05}
This is very much different from La$_{2-x}$Sr$_x$CuO$_4$ where holes
are localized and hence the compound is an 
Anderson insulator~\cite{Boebinger96,ando02} at
$x \lesssim 0.15$.~\cite{com1}
Ultimately, at the very low doping, $x\lesssim 0.01$, the disorder
wins even in YBCO and it also becomes the Anderson insulator.~\cite{WO01}
It is helpful to have in mind an approximate empiric formula~\cite{WO01,Liang}
\begin{equation}
\label{xy}
x\approx 0.35(y-6.20)
\end{equation}
to relate the doping level $x$ and the oxygen content $y$ in 
underdoped YBCO at $x \lesssim 0.12$.

\begin{figure}[ht]
\hspace{-22pt}\includegraphics[width=0.37\textwidth,clip]{fig1top.eps}
\vspace{10pt}
\includegraphics[width=0.4\textwidth,clip]{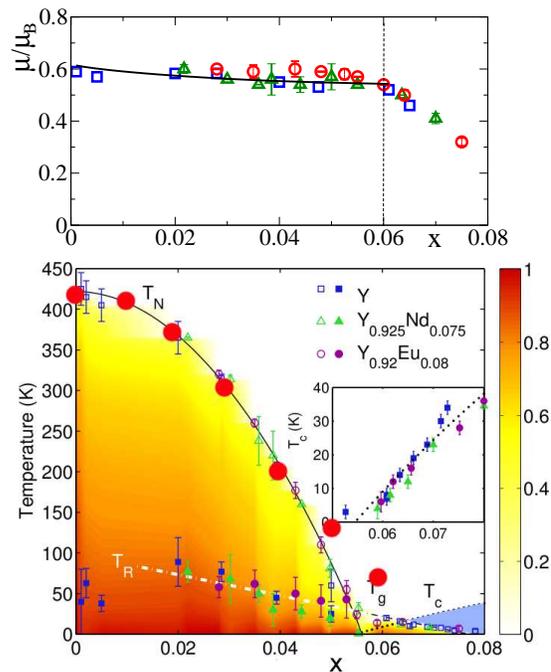}
\caption{\emph{(Color online).
The YBCO $\mu$SR data from Ref.~\onlinecite{con10}.
{\bf Top}:  Zero temperature ``staggered'' magnetization
versus doping. The solid line shows results of the 
present calculations.
{\bf Bottom}: The Neel temperature and the staggered magnetization
versus doping. The inset shows the superconducting critical
temperature versus doping. Large red dots show the Neel temperature
calculated in the present work.
}}
\label{MSR1}
\end{figure}

\begin{figure}[ht]
\includegraphics[width=0.3\textwidth,clip]{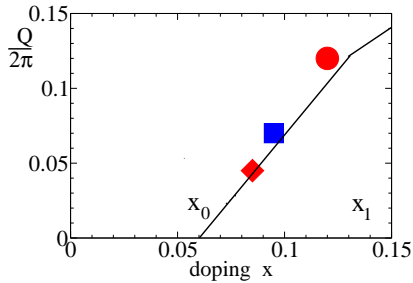}
\caption{\emph{(Color online).
Incommensurate wave vector versus doping.
The blue square~\cite{Stock04},
the red circle~\cite{hinkov07},
and the red diamond~\cite{hinkov08}
show neutron scattering data.
The solid line shows the theoretical value.~\cite{sushkov09}
}}
\label{QQ1}
\end{figure}

The static ``staggered'' magnetization in YBCO has been recently measured in
the $\mu$SR experiment.~\cite{con10}
The experimental plot of the zero temperature magnetization versus doping is 
shown in Fig.~\ref{MSR1}(top).
Remarkably the zero temperature magnetization is almost doping independent 
up to $x \approx 0.06$ and then it quickly decays. It is known from  neutron 
scattering experiments~\cite{hinkov07,hinkov08,hinkov04} that
the static magnetization fully disappears at the
Quantum Critical Point (QCP)  $x\approx 0.09$ indicating
transition to a state without static magnetism.
Importantly, the  magnetism at $x > 0.06$ is incommensurate, this is
why in the first sentence of this paragraph I put ``staggered'' in 
inverted commas.
Value of the incommensurate wave vector $Q$ divided by $2\pi$
versus doping is plotted in Fig.~\ref{QQ1}.

While in the collinear antiferromagnetic phase the zero temperature
staggered magnetization is almost independent of doping, the Neel temperature
 decays very quickly from $T_N=420K$ at $x=0$ to
practically zero at  $x \approx 6\%$.
This is shown in Fig.~\ref{MSR1}(bottom) copied from Ref.~\onlinecite{con10}.
The present paper explains these puzzling magnetic properties and
shows that they are related to small hole pockets of lightly doped Mott
insulator.

YBa$_2$Cu$_3$O$\rm _{y}$ is doped via filling  oxygen chains
located above the CuO$_2$ planes.
It has been argued that at $y=6.5$, where the every second chain
is full, the chain modulation causes the corresponding 
charge density wave (CDW) of in-plane holes.~\cite{Ymany04,feng04}
Nuclear quadruple resonance (NQR) for in-plane Cu is an excellent local
probe   of the hole density.~\cite{Haase04}
\begin{figure}[ht]
\hspace{-5pt}\includegraphics[width=0.5\textwidth,clip]{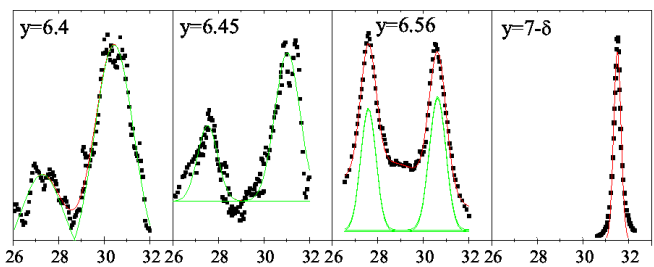}
\caption{\emph{(Color online).
In-plane $^{63}$Cu NQR frequency sweeps from Ref.~\onlinecite{ofer09}.
}}
\label{NQROK}
\end{figure}
Fig.\ref{NQROK} shows $^{63}$Cu NQR frequency sweeps from 
Ref.~\onlinecite{ofer09} for several values of oxygen content.
There is a single narrow line at about optimal doping $y\approx 7$
indicating a  very homogeneous hole density corresponding to
completely filled chains.
On the other hand, at $y\approx 6.5$ there are two distinct lines
indicating a bimodal hole density distribution
in agreement with Refs.~\onlinecite{Ymany04,feng04}.
Importantly, the bimodal distribution is evident even at lower
doping, $y=6,4,6.45$,  indicating the CDW induced by the oxygen chain
superstructure.
Below $y=6.5$ the NQR lines are broader compared to
$y=6.5$. This is because the oxygen superstructure with every second chain
filled cannot be perfect away from $y=6.5$.
It is worth noting that holes go to the CuO$_2$ plane only at $y > 6.2$,
see Eq.(\ref{xy}).
In  the undoped case, $x=0$, $y < 6.2$,  there is
only one NQR line with frequency $\nu_0 \approx 23.3MHz$ independent 
of $y$, see Ref.~\onlinecite{mendels90}.
Comparing $\nu_0$ with frequencies of lines in Fig.\ref{NQROK} we see that 
the hole doping
shifts the NQR frequency very strongly.~\cite{Haase04}
The present paper explains significance of the chain induced CDW
for stability of the collinear antiferromagnetic phase at $x < 0.06$.

Structure of the paper is the following.
The effective theory describing YBCO at low doping was formulated
before in Ref.~\onlinecite{sushkov09}. Section II summarizes ideas of the effective theory.
In Section III the theory is applied to calculate  reduction of the staggered
magnetization in the antiferromagnetic phase at zero temperature.
Temperature reduction of the staggered magnetization at zero and nonzero
doping, $0< x < 0.06$, is calculated in Section IV. 
Interplay between the chain induced CDW, small hole pockets, and
stability of the collinear antiferromagnetic phase is discussed
in Section V. Section VI presents conclusions of the paper.

\section{Effective low energy theory describing lightly doped YBCO} 
This section summarizes the most important points of the effective
low energy suggested in Refs.\onlinecite{milstein08,sushkov09} to describe YBCO
at low doping.
The analysis  is based on  the two-dimensional  $t-t'-t''-J$ model at small 
doping. The generic case of the single layer has been considered in
Ref.~\onlinecite{milstein08}.
After integrating out the high energy fluctuations one comes to the
effective low energy action of the model.
The effective low-energy Lagrangian is written in terms of 
the bosonic ${\vec n}$-field ($n^2=1$) that describes the staggered \
component of the  copper spins,
and in terms of fermionic holons $\psi$. The term ``holon'' is used 
instead of ``hole'' to stress that
 spin and charge are to large extent separated, see 
Ref.~\onlinecite{milstein08}.
The holon has a pseudospin that originates from two sublattices,
so the fermionic field $\psi$ is the spinor in the pseudospin space.
Minimums of the holon dispersion are at the nodal points
$\mathbf{q}_{0}=(\pm \pi /2,\pm \pi /2)$.
So, there are holons of two types corresponding to two
pockets. The dispersion in a pocket is somewhat
anisotropic, but for simplicity let us use here the isotropic approximation,
$
\epsilon \left( \mathbf{p}\right) \approx \frac{1}{2}\beta \mathbf{p}^{2}$
\ ,
where ${\bf p}={\bf q}-\mathbf{q}_{0}$.
The lattice spacing is set to be equal to unity, 3.81\thinspace \AA $\,\rightarrow $
\thinspace 1. 
All in all, the effective Lagrangian for the single layer 
reads~\cite{milstein08}
\begin{eqnarray} 
\label{eq:LL1}
{\cal L}&=&\frac{\chi_{\perp}}{2}{\dot{\vec n}}^2-
\frac{\rho_s}{2}\left({\bm \nabla}{\vec n}\right)^2\\
&+&\sum_{\alpha}\left\{ \frac{i}{2}
\left[\psi^{\dag}_{\alpha}{{\cal D}_t \psi}_{\alpha}-
{({\cal D}_t \psi_{\alpha})}^{\dag}\psi_{\alpha}\right]\right.\nonumber\\
&-&\left.\psi^{\dag}_{\alpha}\epsilon({\bf \cal P})\psi_{\alpha}  
+ \sqrt{2}g (\psi^{\dag}_{\alpha}{\vec \sigma}\psi_{\alpha})
\cdot\left[{\vec n} \times ({\bm e}_{\alpha}\cdot{\bm \nabla}){\vec n}\right]\right\} \ .
\nonumber
\end{eqnarray}
The first two terms in the Lagrangian represent the usual nonlinear 
$\sigma$ model.
The magnetic susceptibility and the spin stiffness are
$\chi_{\perp}\approx 0.53/8\approx 0.066$ and $\rho_s \approx 0.175$~\cite{SZ}.
Hereafter the antiferromagnetic exchange of the initial t-J model is set to be equal to unity, 
\begin{eqnarray}
J\approx 130\thinspace \mbox{meV} \,\rightarrow 
\thinspace 1. \nonumber
\end{eqnarray}
Note that $\rho_s$ is the bare spin stiffness, therefore by definition it is
independent of doping. 
The rest of the Lagrangian in Eq.~(\ref{eq:LL1}) represents the fermionic holon 
field and its interaction with the ${\vec n}$-field.
The index $\alpha=a,b$  indicates the pocket
in which the holon resides.
The pseudospin operator is $\frac{1}{2}{\vec \sigma}$,  and 
${\bf e}_{\alpha}=(\pm 1/\sqrt{2},1/\sqrt{2})$ is a unit  vector orthogonal to 
the face  of the magnetic Brillouine zone (MBZ), where the holon is located. 
The argument of $\epsilon_{\alpha}$ in  Eq.~(\ref{eq:LL1})  
and the time derivative of the  fermionic field in the same equation
are  ``long'' (covariant) derivatives,
\begin{eqnarray}
{\bf {\cal P}}&=&-i{\bm \nabla}
+\frac{1}{2}{\vec \sigma}\cdot[{\vec n}\times{\bm \nabla}{\vec n}] \nonumber\\
{\cal D}_t&=&\partial_t
+\frac{i}{2}{\vec \sigma}\cdot[{\vec n}\times{\dot{\vec n}}] \ .\nonumber
\end{eqnarray}
The covariant derivatives reflect gauge invariance of the initial
$t-t'-t''-J$ model.

Numerical calculations within the $t-t'-t''-J$ model with physical values
of hopping matrix elements give the following values of the coupling constant
and the inverse mass, $g\approx 1$, $\beta \approx 2.4$.
The value of the inverse mass $\beta=2.4$  corresponds to the effective mass
$m^*=1.8m_e$.
The dimensionless parameter
\begin{equation}
\label{Omega}
\lambda=\frac{2g^2}{\pi\beta\rho_s}
\end{equation}
plays the defining role in the theory.
If $\lambda \leq 1$, the ground state corresponding to the Lagrangian 
(\ref{eq:LL1})
is the usual N\'eel state, the state is  collinear at  any small doping.
If $1\leq \lambda \leq 2$, the N\'eel state is unstable at arbitrarily small doping
and the ground state is a static or a dynamic spin spiral.
The wave vector of the spiral is
\begin{equation}
\label{Q1}
Q=\frac{g}{\rho_s}x  \ .
\end{equation}
If $\lambda \geq 2$, the system is unstable with respect to phase separation \
and/or charge-density-wave formation and
hence the effective long-wave-length Lagrangian (\ref{eq:LL1}) becomes 
meaningless.
The pure $t-J$ model ($t'=t''=0$) is unstable since it corresponds
to $\lambda > 2$.

To find parameters of the effective action (\ref{eq:LL1})
one can rely on calculations within the $t-t'-t''-J$ model
or alternatively one can fit experimental data. Both approaches produce
very close values of the parameters.
The fit of elastic and inelastic neutron  scattering 
data for La$_{2-x}$Sr$_x$CuO$_4$ performed in Ref.~\onlinecite{milstein08}
gives the following values, 
$g=1$, $\beta \approx 2.7$ ($m^*=1.5m_e$), $\lambda \approx 1.30$.
The fit of data on magnetic quantum oscillations in 
YBa$_{2}$Cu$_{3}$O$_{y}$ performed in Ref.~\onlinecite{wei10} 
gives two possible sets,
\begin{eqnarray}
\label{gbl}
&&g=1, \ \ \beta = 2.78 \ (m^*=1.45m_e), \ \ \lambda = 1.31,\nonumber\\
&&g=1, \ \ \beta = 2.95 \ (m^*=1.35m_e), \ \ \lambda = 1.23 \ .
\end{eqnarray}
These values will be used in the present work.

It is very easy to understand the reason for instability of the 
commensurate AF order
under doping. Assuming such an order one can calculate the magnon Green's
function
\begin{equation}
\label{gaf}
G(\omega,{\bm q})\propto \frac{1}{\omega^2-c^2q^2-{\cal P}(\omega,{\bm q})+i0}
\ ,
\end{equation}
where $c=\sqrt{\rho_s/\chi_{\perp}}\approx 1.17\sqrt{2}J$ is the magnon 
speed in the parent Mott insulator  and ${\cal P}(\omega,{\bm q})$
is fermionic polarization operator. A well known peculiarity of the
two-dimensional (2D) polarization operator is its independence of doping
as soon as $\omega=0$ and $q$ is sufficiently small.
A straightforward calculation gives at $q\to 0$,
${\cal P}(0,{\bm q})=-\lambda c^2q^2$.
Hence, at $\lambda > 1$ the Stoner criterion in (\ref{gaf}) is violated,
the Green's function possesses poles at imaginary frequency indicating 
instability of the AF ground state at an arbitrary small doping.

\begin{figure}[ht]
\includegraphics[width=0.4\textwidth,clip]{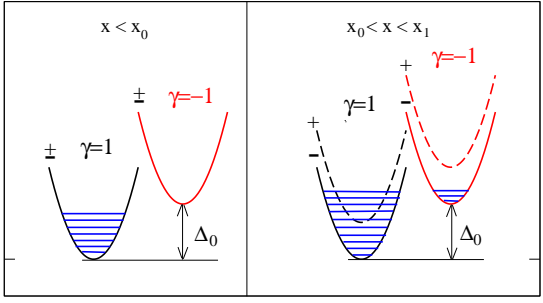}
\caption{\emph{(Color online).
Filling of split holon bands in YBCO at $x < x_0$ (left) and 
$x > x_0$ (right).
The solid and the dashed lines in the spin-spiral state, $x > x_0$,
correspond to different pseudospin
projections, the splitting is $\pm gQ$. 
The doping $x_1$ indicated on the top of the right figure is
$x_1=0.5x_0/(\lambda-1)$, see Ref.~\onlinecite{sushkov09}.
}}
\label{DD2}
\end{figure}

In YBCO the AF order is commensurate at $x < 0.06$, therefore
 the effective action  (\ref{eq:LL1}) cannot be 
directly applied to this compound.
To understand YBCO one can certainly assume that $\lambda$ is doping dependent,
$\lambda < 1$ at $x < 0.06$ and $\lambda > 1$ at $x > 0.06$.
Purely theoretically it is hardly possible to have a significant $x$-dependence
of $\lambda$,  but as a scenario one can consider this.
However, in this scenario the incommensurate wave vector $Q$ must 
jump from $Q=0$ at $x < 0.06$ to $Q$ given by Eq.(\ref{Q1}) at $x > 0.06$.
This is not consistent with data, there is no a jump,
the incommensurate wave vector evolves smoothly above $x=0.06$,
see Fig.\ref{QQ1}, 

A model describing the smooth evolution of $Q$ with doping was
suggested in Ref.~\onlinecite{sushkov09}.
In addition to (\ref{eq:LL1}) the model incorporates two points.
(i) Due to the bilayer structure the magnon spectrum in YBCO is 
split into acoustic and optic mode.
The optic gap is about 70meV,~\cite{Jp} hence the optic mode
does not influence the low energy dynamics, only acoustic magnons
are important for these dynamics. 
(ii) The second point of the model is an assumption that the
fermionic dispersion is split in two brunches as it is shown
in Fig~\ref{DD2}(left). The splitting is $\Delta_0$.
The effective action that originates from (\ref{eq:LL1}) and incorporates
these two points reads
\begin{eqnarray}
\label{eq:LL2}
{\cal L}&=&2\times\left[\frac{\chi_{\perp}}{2}{\dot{\vec n}}^2-
\frac{\rho_s}{2}\left({\bf \nabla}{\vec n}\right)^2\right]\nonumber\\
&+&\sum_{\alpha=a,b}\sum_{\gamma=\pm 1}\left\{ \frac{i}{2}
\left[\psi^{\dag}_{\alpha,\gamma}{{\cal D}_t \psi}_{\alpha,\gamma}-
{({\cal D}_t \psi_{\alpha,\gamma})}^{\dag}\psi_{\alpha,\gamma}\right]\right.\nonumber\\
&-&\psi^{\dag}_{\alpha,\gamma}\left[\epsilon_{\alpha}({\bf\cal P})-\gamma\frac{\Delta_0}{2}\right]\psi_{\alpha,\gamma}\nonumber\\
&+&\left. \sqrt{2}g (\psi^{\dag}_{\alpha,\gamma}{\vec \sigma}\psi_{\alpha,\gamma})
\cdot\left[{\vec n} \times ({ \bf e}_{\alpha}\cdot{ \bf \nabla}){\vec n}\right]\right\} \ .
\end{eqnarray}
Compared to (\ref{eq:LL1}) the first line is multiplied by  two
since the bilayer has the twice larger spin stiffness and magnetic 
susceptibility.
In addition to the pocket index $\alpha$ the holon field
$\psi_{\alpha,\gamma}$ gets an additional index $\gamma=\pm 1$
that indicates the branch of the split dispersion as it is shown in
Fig~\ref{DD2}.
Originally the paper~\cite{sushkov09} suggested that the hole band splitting
$\gamma=\pm 1$ was due to the hole hopping between  layers inside
the bilayer. So, $\Delta_0$  was the bonding-antibonding splitting.
However, our recent analysis~\cite{wei11} indicates that 
antiferromagnetic  correlations forbid the bonding-antibonding splitting.
So, contrary to the assumption in Ref.~\onlinecite{sushkov09} the interlayer
hopping cannot contribute to $\Delta_0$.
In Section VI of the present paper I  argue that the splitting 
$\Delta_0$ is due to oxygen chains. For now let us accept the action
 (\ref{eq:LL2}) and study consequences of this action.

When doping is sufficiently small,
\begin{equation}
\label{x0}
x < x_0=\frac{\Delta_0}{\pi\beta} \ ,
\end{equation}
only the $\gamma=1$ band is populated, see Fig.~\ref{DD2}(left).
In this case the fermionic polarization operator is a half
of that for the single layer case, 
${\cal P}(0,{\bm q})=-\frac{1}{2}\lambda c^2q^2$, $q\to 0$.
Hence the Stoner stability criterion in Eq. (\ref{gaf}) is fulfilled
and the Neel order is stable. 
According to both neutron scattering data~\cite{Stock04,hinkov07,hinkov08}
shown in Fig.~\ref{QQ1} and to $\mu$SR data~\cite{con10} shown in 
Fig.~\ref{MSR1}
the value of $x_0$ is $x_0\approx 0.06$.  Hence, due to Eq.(\ref{x0})
the band splitting is 
\begin{equation}
\label{d0}
\Delta_0\approx 0.5J \approx 65meV \ .
\end{equation}
At $x > x_0$  fermions  populate both $\gamma=1$ and $\gamma=-1$ bands,
Fig.~\ref{DD2}(right), the polarization operator is  doubled compared to the
$x < x_0$ case, and the Stoner instability is there.
As a result at $x > x_0$ the system develops the spiral with the wave 
vector~\cite{sushkov09}
\begin{equation}
\label{Q2n}
Q=\frac{g}{\rho_s}\frac{x-x_0}{3-2\lambda}  \ .
\end{equation}
The plot of $Q/2\pi$ versus doping is shown in Fig.~\ref{QQ1} by the
solid line. The development of the spin spiral is driven by
the pseudospin splitting of the fermionic bands $\pm gQ$
as it is shown  by solid and dashed lines in Fig.~\ref{DD2}(right).
Thus, $x_0$ is a Lifshitz point, where the $\gamma=-1$ band starts to populate,
and where simultaneously the spin spiral starts to develop.
In the present paper I consider quantum and thermal fluctuations
in the Neel state, $x < x_0$.
Quantum fluctuations in the spin spiral state at $x > x_0$ will be considered 
separately.~\cite{milstein11}

To summarize this section.
Small hole pockets and associated  spin spiral state are
generic properties of all cuprates at low doping.
The key point of the YBCO phenomenology  is splitting of the hole pockets. 
This splitting together with splitting of magnon to the acoustic and
the optic mode provides stability of the AF order up to 6\% doping.

\section{Quantum fluctuations in the Neel phase, reduction of the 
zero temperature staggered magnetization.}

There are two magnons (two polarizations) in the Neel phase at $ x < x_0$.
The Green's function of each magnon reads
\begin{equation}
\label{gaf1}
G(\omega,{\bm q})= \frac{(2\chi_{\perp})^{-1}}
{\omega^2-c^2q^2-{\cal P}_0(\omega,{\bm q})+i0}\ .
\end{equation}
Only the $\gamma=1$ band, see Fig.~\ref{DD2}(left), contributes
to the polarization operator ${\cal P}_0(\omega,{\bm q})$.
 Calculation of the polarization operator for 
the single layer was performed   in Ref.~\onlinecite{milstein08}.
Comparing the single layer action (\ref{eq:LL1}) and the double layer
action (\ref{eq:LL2}) and having in mind that at
$ x < x_0$ only the $\gamma=1$ band is occupied one immediately concludes
that in the double layer case the polarization operator is a half of that
calculated in Ref.~\onlinecite{milstein08}. Hence
\begin{widetext}
\begin{eqnarray}
\label{pol0}
{\text {Re}}\ {\cal P}_0(\omega,q)&=&-\frac{c^2g^2}{\pi\beta^2\rho_s}\left\{\beta q^2
-R_1\sqrt{1-R_0^2/R_1^2}\ \ \theta(1-R_0^2/R_1^2)
-R_2\sqrt{1-R_0^2/R_2^2}\ \ \theta(1-R_0^2/R_2^2)\right\}\ , \nonumber\\
{\text {Im}}\ {\cal P}_0(\omega,q)&=&-\frac{c^2g^2}{\pi\beta^2\rho_s}
\left\{ \theta(R_0^2-R_1^2)\ \sqrt{R_0^2-R_1^2}-
\sqrt{R_0^2-R_2^2}\ \theta(R_0^2-R_2^2)\right\}\ , \nonumber\\R_0&=&\beta q p_F \ , \ \ \ R_1=\frac{1}{2}\beta q^2-\omega \ , \ \ \ 
R_2=\frac{1}{2}\beta q^2+\omega \ , \ \ \ p_F=\sqrt{2\pi x} \ .
\end{eqnarray}
\end{widetext}
Here $p_F$ is the Fermi momentum  of the $\gamma=1$ band
 and $\theta(x)$ is the usual step function.
I've already pointed out above that at $q < 2p_F$,
${\cal P}_0(0,{\bm q})=-\frac{\lambda}{2} c^2q^2$,
so the Neel state is stable if $\lambda < 2$.

\begin{figure}[ht]
\hspace{-5pt}\includegraphics[width=0.42\textwidth,clip]{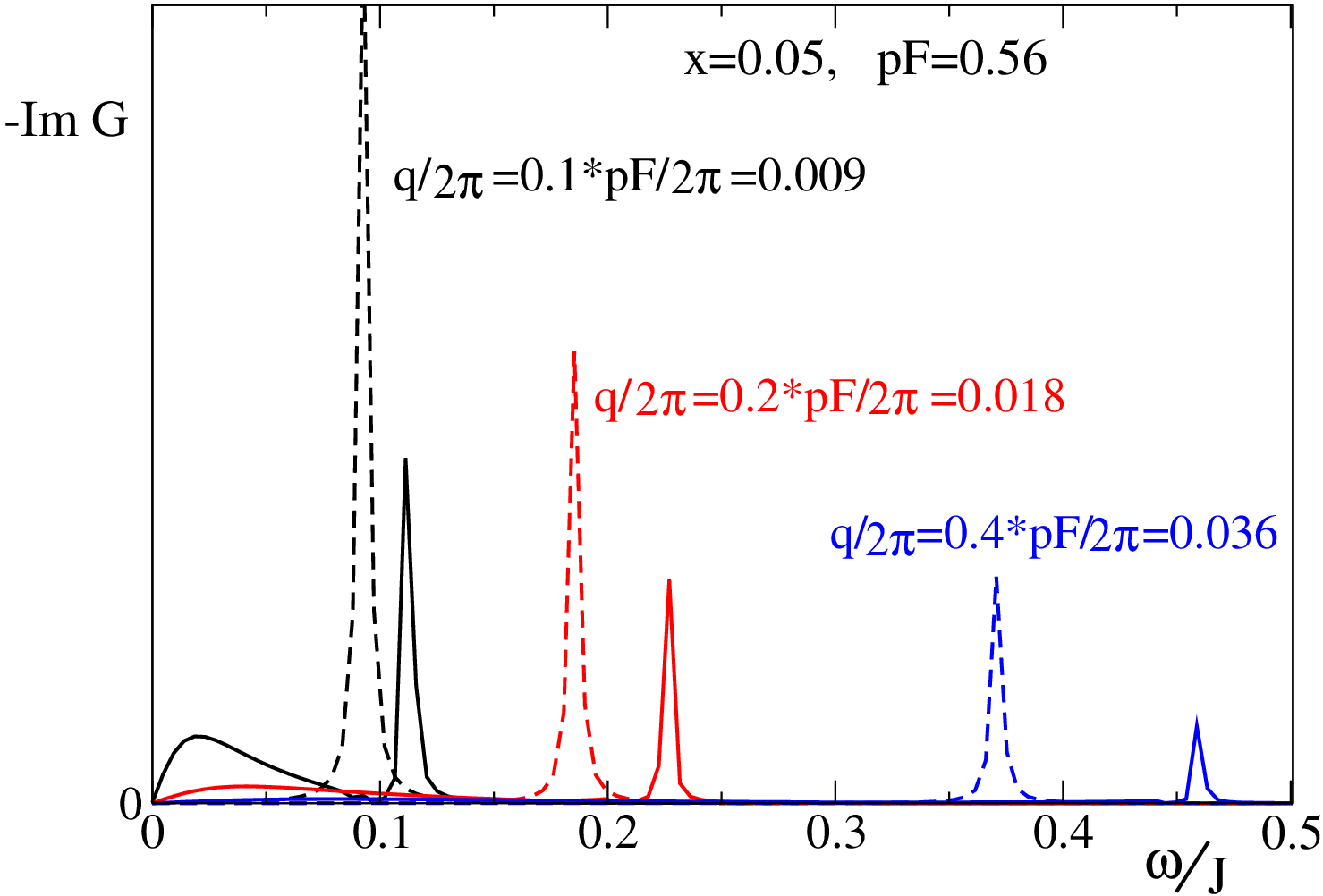}
\caption{\emph{(Color online).
Solid lines show magnon spectral functions $-Im G(\omega, q)$ versus 
$\omega$ for three values of momentum $q$ and for doping $x=0.05$.
Dashed lines show spectral functions in the parent Mott insulator
at the same values of momentum.
}}
\label{ImG}
\end{figure}

It is instructive to look at the magnon spectral function $-Im G(\omega, q)$
that describes inelastic neutron scattering.
Spectral functions for $x=0.05$  and for three values of the momentum $q$
are plotted in Fig.~\ref{ImG} by solid lines.
Spectral functions for both sets of parameters from Eq. (\ref{gbl})
are almost identical. To be specific I present functions for the second set.
In the same Fig.~\ref{ImG}  the dashed lines show spectral functions
in the parent Mott insulator [${\cal P}_0(\omega,{\bm q})=0$].
The spectra demonstrate the low energy incoherent part that absorbs more that
50\% of the spectral weight. The magnon quasiparticle peaks are
still clearly pronounced. Their intensities are about half of that
in the parent compound, and positions are slightly shifted up
compared to the parent, the shift is proportional to the doping,
$\delta \omega_q \propto x$. 
It is worth noting that while the reduction of the spectral weight
is a reliable result, the upward shift is   probably a byproduct of the
low energy effective theory. The effective theory accurately accounts
for the magnon ``repulsion'' from the particle-hole continuum
that is below the magnon.
The ``repulsion''  results in the upward shift.
However, there is also a ``repulsion'' from very high energy excitations
($E \sim 2t \sim 6J$) that are related to the incoherent part of the hole
Green's function. This repulsion, unaccounted in the effective theory,
 leads to the downward shift of the magnon frequency that is also proportional 
to  doping.~\cite{Khaliullin}
More generally one can say that the chiral effective field theory
employed in the present work allows controllable calculations of effects that
are $x$-independent or scale as $\sqrt{x}$ or $x\ln(x)$.
Quantities that scale as the first or as a higher than first power 
of $x$  are generally beyond the scope 
of the theory.
Therefore, at this stage one can say only that position of the magnon
is approximately the same as that in the parent compound, but the
magnon spectral weight is significantly reduced.
Another point worth noting is the absence of the hourglass
dispersion. The low energy incoherent part of the Green's function 
clearly pronounced  in
Fig.~\ref{ImG} is transformed to the coherent hourglass only at $x > x_0$,
beyond the Lifshitz  point.~\cite{milstein11}

Quantum fluctuation of the staggered magnetization is given by the
standard formula
\begin{eqnarray}
\label{n2}
\langle n_{\perp}^2\rangle=-2\sum_{\bf q}\int\frac{d\omega}{2\pi i}
G(\omega,{\bf q})=-2\sum_{\bf q}\int\frac{d\omega}{2\pi}
Im G(\omega,{\bf q})\nonumber\\
\end{eqnarray}
The factor 2 comes from two polarizations.
This expression must be renormalized by subtraction of the 
ultraviolet-divergent contribution that corresponds to the undoped 
$\sigma$-model.
The integral in (\ref{n2}) can be calculated analytically with
logarithmic accuracy
\begin{eqnarray}
\label{n21}
\langle n_{\perp}^2\rangle\approx \frac{\lambda \beta x}{4\rho_s}
\ln\left(\frac{\Lambda}{p_F}\right)
=\frac{\lambda \beta x}{8\rho_s}
\ln\left(\frac{\Lambda^2}{\pi\beta x}\right) \ .
\end{eqnarray}
There are two points to note. (i) In spite of the ultraviolet renormalization
($\sigma$-model subtraction) the fluctuation depends on the ultraviolet
momentum cutoff $\Lambda \sim 1$. (ii) The leading logarithmic term, 
$x\ln(\Lambda^2/x)$, comes from momenta $p_F \ll q \ll \Lambda$.

\begin{figure}[ht]
\hspace{-5pt}\includegraphics[width=0.3\textwidth,clip]{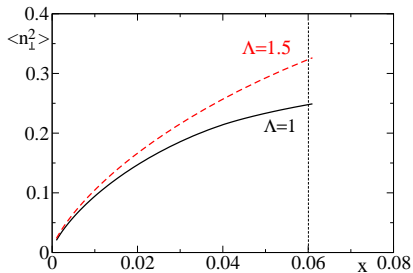}
\caption{\emph{(Color online).
Quantum fluctuation $\langle n_{\perp}^2\rangle$  versus doping
for two values of the ultraviolet cutoff $\Lambda$.
}}
\label{prez}
\end{figure}

The logarithm $\ln(\Lambda^2/x)$ is not large, the logarithmic accuracy 
is not sufficient. Fortunately a numerical integration of (\ref{n2}) is 
straightforward. The result is presented in Fig.~\ref{prez}, where
 $\langle n_{\perp}^2\rangle$  is plotted versus doping.
The second set of parameters from (\ref{gbl}) is used, results are
presented for two values of the ultraviolet cutoff $\Lambda$.
Reduction of the static component of the n-field is
$\langle n \rangle=\langle \sqrt{1-n_{\perp}^2} \rangle 
\approx 1-\frac{1}{2}\langle n_{\perp}^2\rangle$.
Hence the staggered magnetization is
\begin{equation}
\label{mred}
\mu/\mu_B=0.615\left(1-\frac{1}{2}\langle n_{\perp}^2\rangle\right) \ .
\end{equation}
Here I take into account that the used regularization procedure
corresponds to the normalization of the  static component of the  n-field
to unity at zero doping when the staggered magnetization is 
$0.615\mu_B$.~\cite{SZ}
The plot of the calculated staggered magnetization $\mu$ versus doping 
together with  experimental data~\cite{con10} is presented in 
Fig.~\ref{MSR1}(top). 
Dependence of the theoretical curve  on $\Lambda$ is pretty weak,
to be specific the curve corresponding to $\Lambda=1$ is presented.
Agreement between the theory and the experiment in the Neel phase is excellent.
Thus, it is understood why quantum fluctuations only slightly reduce the
staggered magnetization.

Note that the presented calculation is valid only in the Neel phase, $x < 0.06$.
Physics in the spin-spiral phase, $x > 0.06$, is very much different 
because of the appearance of the soft ``hourglass'' dispersion and
consequently because of greatly enhanced quantum fluctuations.
The corresponding results will be published separately.~\cite{milstein11}

\section{Temperature dependence of the staggered magnetization in  
the Neel phase}

\subsection{Zero doping}
It is well known that due to the Mermin-Wagner theorem the 
Neel temperature in a spin-rotationally-invariant 2D system is exactly zero,
$T_N=0$.
Cuprates are layered systems with a very small Heisenberg interaction,
$J_{\perp} \lesssim 10^{-4}J$, between layers or bilayers.
In spite of its smallness the interaction makes the system three dimensional
 and hence it makes the Neel temperature finite,
$T_N \sim J/\ln(J/J_{\perp})$.
Temperature dependence of the staggered magnetization
in layered Heisenberg antiferromagnets has been
intensively studied theoretically, for a review see Ref.~\onlinecite{Katanin}.
Unfortunately there is no a ``small theoretical parameter'' in the problem,
therefore, while a qualitative behaviour is absolutely clear, there is no
a universal quantitative description,
different theoretical approaches give quite different results.~\cite{Katanin}
In the present section I develop an effective description of the
temperature dependence. This description certainly is not a rigorous solution of
the layered Heisenberg antiferromagnet for all temperatures.
This is a sort of  interpolation between $T \ll T_N$ regime and 
$T\approx T_N$ regime.
Importantly, the ``interpolation'' allows to describe 
quantitatively an undoped layered Mott
insulator, and much more importantly it allows to move to the finite doping
in the next subsection.

\begin{figure}
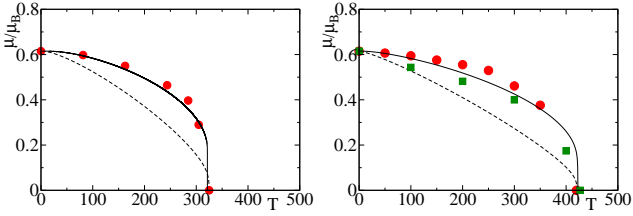

\includegraphics[width=0.23\textwidth,clip]{fig7L.eps}
\includegraphics[width=0.23\textwidth,clip]{fig7R.eps}
\caption{Staggered magnetization versus temperature in 
La$_{2}$CuO$_4$, left, and YBa$_{2}$Cu$_{3}$O$_{6}$, right.
In the left plot red circles show neutron scattering data.~\cite{Keimer92}
In the right plot red circles show neutron scattering data~\cite{RM}
and  green squares show $\mu$SR data.~\cite{con10} 
Theoretical curves with $r=2$ are shown by  
solid lines and theoretical curves with $r=1$ are shown by  
dashed lines.
 }
\label{lym}
\end{figure}

Let us start from the single layer case (La$_{2}$CuO$_4$) and rewrite
Eq.(\ref{n2}) in the Matsubara technique  at a finite temperature.
\begin{eqnarray}
\label{n2T}
\langle n_{\perp}^2\rangle=\frac{2T}{\chi_{\perp}}
\sum_{\bf q}\sum_s\frac{1}{\xi_s^2+\omega_q^2} \ ,
\end{eqnarray}
where $\omega_q=cq$ and $\xi_s=2s\pi T$, $s=0,\pm 1,\pm 2, ...$ is the Matsubara
frequency.
Hence equation for $n_z=1-\frac{1}{2}\langle n_{\perp}^2\rangle$
can be rewritten in the renormalization group (RG) form
\begin{eqnarray}
\label{gnz}
\frac{dn_z}{n_zd\ln(q)}=\frac{T}{2\pi\rho_{sq}}\sum_s\frac{\omega_q^2}{\omega_q^2+\xi_s^2} \ ,
\end{eqnarray}
where $\rho_{sq}=\rho_s(q)$ is the $q$-dependent spin stiffness.
Eq.(\ref{gnz}) assumes 2D geometry, so it is valid at
$q > q_{min}$, where the infrared cutoff $q_{min}\propto \sqrt{J_{\perp}}$  
is due to the Heisenberg coupling along the third dimension.
To solve the RG problem one needs to add information how the spin stiffness
scales with the magnetization.
Let us write the relation between the magnetization and the spin stiffness as
\begin{eqnarray}
\label{rnz}
\frac{d\rho_{sq}}{\rho_{sq}}=r\frac{dn_z}{n_z} \ . 
\end{eqnarray}
It is known~\cite{Katanin} that one loop calculation valid at $n_z \approx 1$
results in $r=1$ that implies  $\rho \propto n_z$.
On the other hand, close to the Neel temperature when $n_z \ll 1$ one
should expect scaling very close to quadratic,
$\rho \propto n_z^2$ ($r\approx 2$). This is because the critical index 
$\eta$ of the magnon quasiparticle residue is very small, see, e.g. 
Refs.~\onlinecite{Chub94,Irkhin97}.
For now I keep the power $r$ as a parameter.
Eqs. (\ref{gnz}) and (\ref{rnz}) are combined to
\begin{eqnarray}
\label{rrnz}
\frac{d\rho_{sq}}{d\ln(q)}=\frac{rT}{2\pi}\sum_s\frac{\omega_q^2}{\omega_q^2+\xi_s^2} 
\end{eqnarray}
To perform the ultraviolet renormalization let us introduce $\rho_{\Lambda}$,
the spin stiffness at the ultraviolet cutoff.
Then due to Eq.(\ref{rrnz}) the finite temperature spin stiffness at
$q=0$ reads
\begin{eqnarray}
\label{rst}
\rho_{sT}=\rho_{\Lambda}-\frac{rT}{2\pi}\int_{q_{min}}^{\Lambda}
\left\{\sum_s\frac{\omega_q^2}{\omega_q^2+\xi_s^2}\right\}\frac{dq}{q}\ .
\end{eqnarray}
This expression can be renormalized by the condition that at zero 
temperature (more accurately at $T \ll cq_{min}$) the spin
stiffness is equal the standard value $\rho_{s0}\approx 0.175J$ 
corresponding the $\sigma$-model originated from the spin 1/2 Heisenberg model.
After the renormalization Eq.(\ref{rst}) is transformed to
\begin{eqnarray}
\label{rint}
\rho_{sT}=\rho_{s0}-\frac{rT}{2\pi}\int_{q_{min}}^{\infty} 
\left\{\sum_s\frac{\omega_q^2}{\omega_q^2+\xi_s^2}-\frac{\omega_q}{2T}\right\}\frac{dq}{q}\ .
\end{eqnarray}
The 3D  interaction $J_{\perp}$ fixes value of the infrared cutoff $q_{min}$, 
however, one has to remember about scaling of the cutoff with the 
staggered magnetization  $n_z$, see Ref.\cite{Katanin},
\begin{equation}
\label{interp}
q_{min}=q_{min0}\sqrt{n_z} \ ,
\end{equation}
where, due to (\ref{rnz}),
\begin{equation}
\label{sr}
n_z=\left[\frac{\rho_{sT}}{\rho_{s0}}\right]^{1/r} \ .
\end{equation}
Eqs. (\ref{rint}),(\ref{interp}),(\ref{sr}) can be easily integrated 
numerically.
The Neel temperature is determined by zero of the spin stiffness (\ref{rint}).
The infrared cutoff $q_{min0}$ is the only  free parameter in the theory.
Value of the parameter has to be tuned up to reproduce the measured
Neel temperature.
 It has to be clear that $q_{min0}$ originates not only from
$J_{\perp}$, relativistic anisotropies like Dzyaloshinskii-Moria etc, 
also contribute to $q_{min0}$.
Let us recall that due to the used regularization procedure the
staggered magnetization is $\mu=0.615\mu_Bn_z$, where $0.615\mu_B$ is the
the staggered magnetization in the parent Heisenberg model.~\cite{SZ}
Staggered magnetization versus temperature in La$_{2}$CuO$_4$
is presented in Fig.~\ref{lym}(left).
Red circles show neutron scattering data.~\cite{Keimer92}
The theoretical curve with $r=2$ and $q_{min0}=0.024$
is shown by  the solid line and the theoretical curve with $r=1$ 
and $q_{min0}=0.004$ is shown by the dashed line.
The curve with $r=1$ corresponding to
the single loop RG describes the data very poorly.
This illustrates the  known problem of poor accuracy of 
the single loop RG.~\cite{Katanin}
However, the curve with $r=2$ corresponding to the critical scaling
of the spin stiffness describes the data very well.

In the double layer case (YBCO) the coefficient $\frac{rT}{2\pi}$ before the
integral in Eq.(\ref{rint})
has to be replaced by the twice smaller one, $\frac{rT}{4\pi}$.
The point is that the optic magnon in YBCO has a gap 70meV and therefore
it does not contribute to the low energy dynamics. 
Only acoustic magnon is important, hence the effective number of magnons is 
twice smaller compared to LCO.
Neutron scattering data~\cite{RM} for YBa$_{2}$Cu$_{3}$O$_{6}$ are shown
in Fig~\ref{lym}(right) by red dots.
Green squares show $\mu$SR data.~\cite{con10}
The theoretical curve with $r=2$ and  $q_{min0}=0.0085$ 
is shown by  the solid line and the theoretical curve with $r=1$ 
and $q_{min0}=0.0004$ is shown by  the dashed  line.
Again, the curve with $r=1$ is not consistent with the data.
The curve with $r=2$ is quite good.

It is worth noting that for both  LCO and YBCO the
values of the infrared cutoff $q_{min0}$ for $r=1$ 
are unrealistically small reflecting the same difficulty
of single loop RG, see also.~\cite{Katanin}
 On the other hand, the cutoff values for $r=2$
are quite reasonable, they indicate that the Neel temperature is
determined by spin-wave dynamics at distances up to $1/q_{min0}\sim 100$
lattice spacing along the plane.

All in all, the conclusion is that the effective RG developed in
this subsection describes undoped compounds pretty well.
To achieve this description one needs to set $r=2$,
this corresponds to the critical scaling of the spin stiffness 
expected in the vicinity of the Neel temperature, $\rho \propto n_z^2$.
 In the next subsection the developed description will be applied
to the nonzero doping case.

\subsection{Nonzero doping}
To extend to the finite doping case one has to introduce in Eq.(\ref{rint}) 
the fermionic polarization operator 
\begin{eqnarray}
\label{rintx}
\rho_{sT}&=&\rho_{s0}\\
&-&\frac{rT}{2\pi}\int_{q_{min}}^{\infty} 
\left\{\sum_s\frac{\omega_q^2}{\omega_q^2+\xi_s^2+{\cal P}_0(i\xi_s,q)}-
\frac{\omega_q}{2T}\right\}\frac{dq}{q}\ ,\nonumber
\end{eqnarray}
where the polarization operator ${\cal P}_0(i\xi_s,q)$ is calculated at 
Matsubara frequencies. 
Expression for the polarization operator follows from the 
Lagrangian (\ref{eq:LL2}). One can use vertexes derived in 
Ref.~\onlinecite{milstein08} for the single layer case and rescale the
vertexes by the factor $1/\sqrt{2}$ that follows from comparison
of (\ref{eq:LL1}) and (\ref{eq:LL2}).
The polarization operator reads 
\begin{eqnarray}
\label{pxi1}
{\cal P}_0(i\xi,{\bm q})&=&\pi\lambda c^2\beta q^2 \mbox{Re}\sum_{\gamma=\pm1}\sum_{\bm p}
\frac{f_{\bm p}^{\gamma}-f_{{\bm p}+{\bm q}}^{\gamma}}{\epsilon_{\bm p}-\epsilon_{{\bm p}+{\bm q}}+i\xi}
\nonumber\\
&=&2\pi\lambda c^2\beta q^2 \mbox{Re}\sum_{\gamma=\pm1}\sum_{\bm p}
\frac{f_{\bm p}^{\gamma}}{\epsilon_{\bm p}-\epsilon_{{\bm p}+{\bm q}}+i\xi} \ .
\end{eqnarray}
Here $f_{\bm p}^{\gamma}$ is the Fermi-Dirac distribution function
\begin{equation}
\label{fd}
f_{\bm p}^{\gamma}=\frac{1}{e^{(\epsilon_{\bm p}-\gamma \Delta_0/2-\mu)/T}+1}
\end{equation}
with chemical potential $\mu$ (do not mix it up with magnetic moment).
Note that at $T\ne 0$ the $\gamma=-1$ band is also populated, 
see Fig.\ref{DD2}(left).
This is why the summation in (\ref{pxi1}) is performed over both bands, $\gamma=\pm 1$.
The chemical potential is determined by the condition
\begin{equation}
\label{chem}
2x=2\times 2 \sum_{\gamma=\pm1}\sum_{\bm p}f_{\bm p}^{\gamma}\ ,
\end{equation}
that accounts for the double layer, the two pockets, and for the two 
psedospin projections.
It is easy to check that at zero temperature and at $q < 2p_F$ the zero
frequency polarization operator is ${\cal P}_0(0,q)=-\frac{\lambda}{2}\omega_q^2$, 
in agreement with the real frequency analysis at $x < x_0$ in section III.

\begin{figure}[ht]
\includegraphics[width=0.3\textwidth,clip]{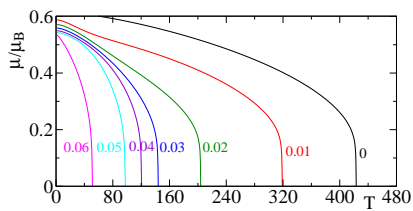}
\caption{Staggered magnetization versus temperature 
in YBa$_{2}$Cu$_{3}$O$_{y}$ for several values of doping.
Values of doping $x=0,0.1,0.2,0.3,0.4,0.5,0.6$
are shown near the corresponding curve.
These curves have been obtained without account of the lifetime of the hole,
$\Gamma=0$.
 }
\label{Yg0}
\end{figure}

\begin{figure}[ht]
\includegraphics[width=0.3\textwidth,clip]{fig9top.eps}
\includegraphics[width=0.32\textwidth,clip]{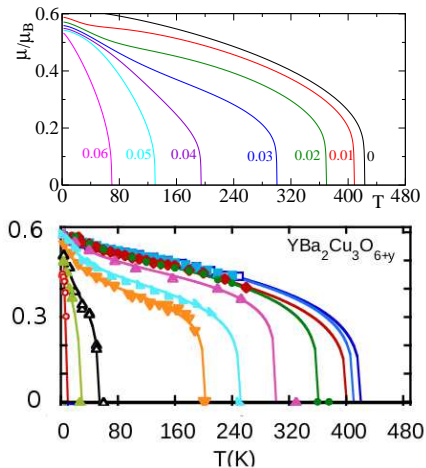}
\caption{Staggered magnetization versus temperature 
in YBa$_{2}$Cu$_{3}$O$_{y}$ for several values of doping $x$.
Theoretical curves with account of the hole lifetime
for $x=0,0.1,0.2,0.3,0.4,0.5,0.6$ are shown in the top figure.
Experimental curves  from Ref.~\onlinecite{con10} are presented in
the bottom figure, the doping levels are
$x=0.001,0.002,0.005,0.02,0.029,0.036,0.039,0.049,0.061,0.065$.
}
\label{Yg}
\end{figure}

Numerical evaluation of the polarization operator (\ref{pxi1}) 
is straightforward. Substitution of the polarization operator in the
RG equation (\ref{rintx}) and solution of this equation together
with  (\ref{interp}) and (\ref{sr}) gives staggered magnetization
at a given doping and temperature.
The RG equation is solved with $r=2$ and $q_{min0}=0.0085$ as it has been
discussed in the previous subsection. These parameters are relevant to
the n-field and they are independent of doping.
Fermionic polarization operator (\ref{pxi1}) is not very sensitive to the
choice of parameters, to be specific I present results corresponding
to the second set of parameters in  Eq. (\ref{gbl}).
The band splitting $\Delta_0$ is determined by Eq.(\ref{d0}) that
is responsible for the position of the Lifshitz point, $x_0=0.06$.
Plots of the staggered magnetization versus temperature for several values of
doping are shown in Fig.~\ref{Yg0}.
Theoretical curves  presented in Fig.~\ref{Yg0} are in a qualitative and to 
some extent quantitative agreement with data from Ref.~\cite{con10}
shown in Fig.~\ref{MSR1}(bottom) and in Fig.~\ref{Yg}(bottom). 
The theory clearly demonstrates that while reductions of the zero temperature
staggered magnetization at $x < x_0$ is pretty small, Fig.~\ref{MSR1}(top),
the reduction of the Neel temperature with doping is dramatic.
There are two reasons for the reduction.
(i) Thermal excitation of the precursor to the hourglass, the
lower incoherent part of the magnetic spectrum shown in Fig.~\ref{ImG}.
(ii) Thermal  population of the $\gamma =-1$ band.  
Fig.~\ref{Yg0} indicates also some negative
bending of $\mu(T)$ curves in a qualitative agreement with data
presented in Fig.~\ref{Yg}(bottom).

Theoretical curves plotted in Fig.~\ref{Yg0} 
demonstrate even too steep decrease of the Neel temperature
with doping  compared to experimental data shown in Fig.~\ref{MSR1}(bottom).
For example at $x=0.03$ the theoretical Neel temperature is 140K, Fig.~\ref{Yg0},
 while experimentally it is about 300K, Fig.~\ref{MSR1}(bottom).
To fix this problem one has to realize that the above consideration of
fermions disregards an important physical effect, the finite lifetime 
(scattering time) of a fermion at a nonzero temperature.
This is the  same effect that leads to the temperature dependent resistivity.  
To understand importance of this effect prior to calculations
one has to recall, 
see previous subsection, that the Neel temperature is formed at very large
in-plane distances up to 100 lattice spacing. 
This corresponds to $q\sim 0.01$ in Eqs.(\ref{rintx}),(\ref{pxi1}).
The fermionic polarization operator at so small $q$ (so large distances)
must depend on the fermion mean free path. This explains crucial
importance of the fermion lifetime.
To account for the lifetime effect Eq.(\ref{pxi1}) has to be modified
in the following way
\begin{eqnarray}
\label{pxi2}
{\cal P}_0(i\xi,{\bm q})
=2\pi\lambda c^2\beta q^2\sum_{\gamma=\pm1}\sum_{\bm p}
\frac{(\epsilon_{\bm p}-\epsilon_{{\bm p}+{\bm q}})f_{\bm p}^{\gamma}}
{(\epsilon_{\bm p}-\epsilon_{{\bm p}+{\bm q}})^2+\xi^2+\frac{\Gamma^2}{4}} \ .
\end{eqnarray}
Here $\Gamma$ is the  broadening due to scattering.
Let us take the usual width characteristic for the two-dimensional 
Fermi liquid~\cite{gia},
\begin{equation}
\label{Gamma}
\Gamma=A\frac{T^2}{\epsilon_F} \ ,
\end{equation}
where $\epsilon_F=\beta p_F^2/2=\pi\beta x$ is the Fermi energy.
I disregard the logarithmic $T$-dependence of the coefficient $A$,
the dependence is beyond accuracy of the calculation.
The coefficient $A$ will be used as a fitting parameter.
Note, that generally the width $\Gamma$ depends on both temperature $T$ and
Matsubara frequency $\xi_s$. The dominating contribution to Eq.(\ref{rintx})
comes from the zero Matsubara frequency. Therefore, the width $\Gamma$
is important in the zero frequency term, $s=0$, and it is completely
negligible in $s\ne 0$ terms. Hence, the width (\ref{Gamma}) corresponds
to the zero Matsubara frequency.
Numerical evaluation of the polarization operator (\ref{pxi2}) is not
more difficult than evaluation of (\ref{pxi1}).
Solution of  RG equations gives the staggered magnetization $\mu(T,x)$
with account of the fermion lifetime. 
The best fit of the experimental
dependence of the Neel temperature on doping is achieved at
\begin{equation}
\label{AA}
A \approx 0.7 \ .
\end{equation}
The calculated Neel temperature versus doping is shown by large
ired dots in Fig.~\ref{MSR1}(bottom).
The calculated staggered magnetization versus temperature
is plotted in Fig.\ref{Yg}(top) for several values of doping.
Experimental curves from Ref.~\onlinecite{con10} are presented in
Fig.~\ref{Yg}(bottom). 
Overall agreement between theory and experiment is very good.

It is worth stressing again that the calculation of the 
temperature dependence of the magnetization
 in the layered system is less reliable than 
calculations of zero temperature properties in section III.
The complexity of the finite temperature case
is due to the very large span of spacial scales
involving in the problem with the largest scale about 100 lattice spacing.
Only leading effects have been taken into account in the present calculation.
Clearly there are subleading effects that also influence the magnetization.
For example, usual disorder (impurities) must influence fermion dynamics
at the scale $\sim 100$ lattice spacing and hence influence magnetization.
In view of this comment the agreement between the theory and
experimental data, see Fig.~\ref{MSR1}(bottom) and Fig.~\ref{Yg},
is remarkable.
Most importantly, the theory explains
why the Neel temperature drops down dramatically with doping, while
the zero temperature magnetization is almost doping independent.
This ``contradictory'' behaviour is due  to the band splitting,
and due to different fillings of the split bands.
The different filling is a fingerprint of small hole pockets.
At zero temperature only the lower band is occupied while temperature
populates the upper band as well. The ``contradictory'' behaviour
 is closely related to the Lifshitz point at $x=x_0$ and to the development
of the spin spiral at $ x > x_0$ when both bands are occupied at zero 
temperature.

 \section{CDW induced by oxygen chains, small hole pockets,  mechanism for
band splitting}

The key point of the YBCO phenomenology suggested in 
Ref.~\onlinecite{sushkov09} and applied in the present work is
the splitting of hole bands. There are other key points like small
hole pockets, spin spirals, etc. However these other points are not
specific to YBCO, they are generic for all cuprates.
The band splitting is specific to YBCO.
The paper~\cite{sushkov09} suggested that the hole band splitting
in YBCO was due to the hole hopping between  layers inside
the bilayer, the bonding-antibonding splitting.
However, our recent analysis~\cite{wei11} indicates that 
antiferromagnetic  correlations {\it between} the layers
forbid the bonding-antibonding splitting.
If the hole hopping matrix element between the layers is $t_{\perp}$
then the band splitting in the case of AF correlations  between the
layers is $\propto t_{\perp}(\cos k_x+\cos k_y)$, see 
Ref.~\onlinecite{wei11}.
The splitting is zero at the nodal points $(k_x,k_y)=(\pm \pi/2,\pm \pi/2)$
contrary to the assumption  (\ref{eq:LL2}).

Thus, contrary to the assumption in Ref.~\onlinecite{sushkov09} the interlayer
hopping cannot contribute to $\Delta_0$. Another mechanism for splitting
is necessary.  In this section I argue that the splitting is due to
oxygen chains.
\begin{figure}
\includegraphics[width=0.2\textwidth,clip]{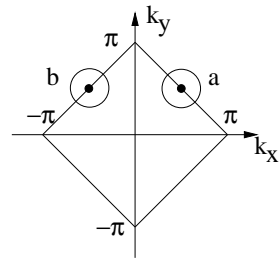}
\caption{Single hole dispersion in the  AF background.
 }
\label{valley}
\end{figure}
Let us first consider the case $y=6.5$ when every second chain is fully filled.
In this case chains produce the effective pseudopotential for in-plane holes
\begin{equation}
\label{vc}
V_c(X)=-v_0\cos(\pi X) \ ,
\end{equation}
where $v_0$ is the amplitude of the potential and
$X$ is the direction perpendicular to chains 
(I denote the distance by capital $X$ to make it different from the doping $x$).
The hole dispersion in the antiferromagnetic background is shown in 
Fig.~\ref{valley}.
The dispersion consists of two full pockets, the pocket $a$ and the pocket $b$.
There is the perfect nesting condition between the pockets and the
chain potential (\ref{vc}), therefore the two split bands are generated
\begin{eqnarray}
\label{dvc}
&&\epsilon_p=-v_0+\frac{\beta}{2}p^2\ , \ 
\psi_+=\frac{|a\rangle_p+|b\rangle_p}{\sqrt{2}} 
\propto \cos\left(\frac{\pi}{2} X\right)e^{i{\bm p}\cdot{\bm r}}\nonumber\\
&&\epsilon_p=+v_0+\frac{\beta}{2}p^2\ , \ 
\psi_-=\frac{|a\rangle_p-|b\rangle_p}{\sqrt{2}}\propto 
\sin\left(\frac{\pi}{2} X\right)e^{i{\bm p}\cdot{\bm r}}\nonumber\\
\end{eqnarray}
Here ${\bm p}$ is the momentum with respect to the center of the pocket.
Eqs. (\ref{dvc}) represent  exactly the $\gamma=\pm 1$ band splitting 
adopted in (\ref{eq:LL2}).
Due to the exact nesting of small hole pockets a tiny pseudopotential 
$v_0\approx 30meV$ is sufficient to generate the splitting
$\Delta_0=2v_0\approx 65meV$ that follows from  the magnetic analysis,
see Eq.(\ref{d0}).

There might be an impression that the splitting (\ref{dvc}) 
is not quite equivalent to the effective action  (\ref{eq:LL2}).
For example, the question arises why there is no a spin-wave vertex
that transfers $\psi_+$ to $\psi_-$? The vertex  carries a large
momentum  $\pi$, therefore the vertex  vanishes after integration over $X$.
In other words, soft magnons included in the effective action (\ref{eq:LL2})
cannot induce a transition with the large momentum transfer.
A careful analysis shows that the splitting (\ref{dvc}) with account 
of two layers is completely equivalent to (\ref{eq:LL2}).

According to (\ref{dvc}) the wave function of the lower $\gamma=+1$ band
is nonzero at X=0,2,4,... while the wave function of the upper 
$\gamma=-1$ band is nonzero at X=1,3,5,...
Due to the splitting the bands are differently populated and this
 results in  the in-plane hole density wave with period of the two lattice 
spacing.~\cite{Ymany04,feng04} 
Let us  calculate the amplitude of the CDW.
The oxygen content $y=6.5$ corresponds to doping $x\approx 0.1$, see
Eq.(\ref{xy}). This doping is within the spin-spiral phase, therefore 
to calculate fillings of bands one has to account the spin spiral
as it is shown in Fig.~\ref{DD2}(right).
In the lower $\gamma=+1$ band
both pseudospin projections are populated,
while in the upper $\gamma=-1$ band
only one pseudispin projection is populated.
Populations of different subbands have been calculated 
in the analysis of magnetic quantum oscillations, see Eqs. (4) in
Ref.~\onlinecite{wei10}.
From these equations one concludes that populations of the  upper and
lower bands ($\gamma=\mp 1$) are
\begin{eqnarray}
\label{xpm}
x_{-1}&=&\frac{2-\lambda}{2(3-2\lambda)}(x-x_0)\nonumber\\
x_{+1}&=&x-x_{-1} \ .
\end{eqnarray}
Naturally, the population of the upper band vanishes at $x=x_0$,
this is the Lifshitz point. For $x=0.1$, $x_0=0.06$, and $\lambda=1.23$,
one finds $x_{-1}=0.03$ and $x_{+1}=0.07$.
Hence, the hole density per cite at every even value of X is $2x_{+1}=0.14$,
and at every odd value of X it is $2x_{-1}=0.06$.
However, this is not the amplitude of the CDW yet.

All equations in the present paper are written in terms
of holes dressed by magnetic quantum fluctuations (magnetic polarons).
Hence $x_{\pm 1}$ are densities of the {\it dressed holes}.
The dressed hole has a finite size, therefore, the charge density
modulation is smaller than that naively given by $x_{\pm 1}$.
It is known that the quasiparticle residue of the dressed hole is about
$Z\approx 0.4$, see e.g. Ref.~\onlinecite{sushkov97}.
This means that with the probability $Z \approx 0.4$ the hole resides at 
the same 
site as the quasihole and with the probability $(1-Z)/4 \approx 0.15$ the hole
resides at each of the four nearest Cu sites. Therefore, the real charge
densities per site are
\begin{eqnarray}
\label{rpm}
\rho_{+1}&=&2\left[\left(Z+2\frac{1-Z}{4}\right)x_{+1}+2\frac{1-Z}{4}x_{-1}\right]
\approx 0.12\nonumber\\
\rho_{-1}&=&2\left[\left(Z+2\frac{1-Z}{4}\right)x_{-1}+2\frac{1-Z}{4}x_{+1}\right]
\approx 0.08\nonumber\\
\end{eqnarray}
This gives the amplitude of the CDW.
The estimate of the amplitude is based purely on magnetic data, it depends 
mainly on the position of the Lifshitz point, $x_0=0.06$.

NQR was not used in the estimate (\ref{rpm}).
Nevertheless  the estimate is pretty much consistent with NQR 
data~\cite{ofer09} presented in Fig.~\ref{NQROK}.
The NQR frequency shift with respect to  the frequency of the
undoped sample, $\nu_0=23.3MHz$,~\cite{mendels90} is proportional to the 
local hole concentration~\cite{Haase04}
\begin{equation}
\label{fs}
\nu_Q=23.3MHz+B\rho \ .
\end{equation}
The higher frequency NQR line at $y=6.56$ is
$\nu_{2}\approx 30.3MHz$, see Fig.~\ref{NQROK}.
According to Eq.(\ref{rpm})  the line corresponds to $\rho\approx 0.12$.
Hence, the constant $B$ in Eq.(\ref{fs}) is $B\approx 58MHz/hole$.
Interestingly, the value of $B$ is significantly larger than that
in La$_{2-x}$Sr$_x$CuO$_4$, $B\approx 20MHz/hole$,~\cite{Haase04} \
and in HgBa$_2$CuO$_{4+\delta}$, $B\approx 30MHz/hole$.~\cite{chen11}
Assuming that optimal doping corresponds to $\rho\approx 0.14$ and 
using Eq.(\ref{fs}) one finds the optimal doping NQR frequency $\nu=31.4MHz$. 
This value is pretty close to  $\nu_{opt}\approx 31.6MHz$ that follows from
Fig.~\ref{NQROK} at $y\approx 7$.
According to (\ref{rpm}) the lower frequency NQR line at $y\approx 6.5$
corresponds to $\rho\approx 0.08$. Substituting this value in Eq.(\ref{fs}) 
one finds the frequency $\nu=27.9MHz$. Again, this value is
pretty close to  the lower frequency line $\nu_{1}\approx 27.8MHz$ that 
is shown in Fig.~\ref{NQROK} at $y=6,56$.
Thus, the amplitude of the CDW determined from the position of the
Lifshitz point is fully consistent with the NQR data.

The simple potential (\ref{vc}) is literally applicable
only to $y=6.5$. Obviously, there is no any modulation at $y=7$
as the rightmost Fig.~\ref{NQROK} indicates.
Away from $y=6.5$ more complex oxygen superstructures can 
appear.~\cite{stempfer}
Assumption important for the present work is that at $0< x<0.1$
($6.20 < y < 6.5$) the superstructure (\ref{vc}) is dominating.
NQR data~\cite{ofer09} for $y=6.4$ and $y=6.45$  
presented in Fig.~\ref{NQROK} support this assumption: there are
only two NQR lines that are only slightly broader than the lines
at $y=6.5$.

\section{Conclusions}
Small hole pockets and the associated  spin spiral state are
generic properties of all cuprates at low doping.
The key point of the YBCO phenomenology  additional to the generic 
properties is splitting of the hole pockets
into the lower band and the upper band.
This splitting together with splitting of magnon to the acoustic and
the optic mode provides stability of the collinear aniferromagnetic
order at doping below the Lifshitz point at $x\approx 0.06$.
At doping below the Lifshitz point only the lower band is
populated.
At higher doping the upper band starts to populate
and simultaneously the spin spiral starts to develop.

At doping below the Lifshitz point the doping induced spin quantum 
fluctuations are pretty weak. This explains why the 
zero temperature staggered magnetization 
is close to $0.6\mu_B$, the maximum value allowed by 
quantum  fluctuations of localized spins.
The developed theory perfectly reproduces the weak decrease of the
staggered magnetization with doping observed experimentally.

While the zero temperature staggered magnetization is almost
doping independent,  the Neel temperature decays very quickly from 
$T_N=420K$ at $x=0$ to practically zero at  $x \approx 0.06$.
This  quick decay is a consequence of the closeness to the
Lifshitz point. 
Again, the theory reproduces very well
the doping dependence of the Neel temperature as well as
the observed temperature dependence of the staggered magnetization
at a given doping.

The band splitting (the hole pocket splitting) is induced by the 
modulation of oxygen chains. The main period  of the modulation is
two lattice spacing. Because of the perfect nesting between 
the small hole pockets and the period of the modulation, a small
pseudopotential caused by the chains is sufficient to induce
the band splitting about 60meV. The splitting causes the 
in-plane charge density wave with  a significant amplitude
dependent on doping.

\acknowledgements
I am grateful to
W. Chen,
J. Haase,
M.-H. Julien, 
A. A. Katanin,
G. Khaliullin,
D. Manske,
W. Metzner,
A.I. Milstein,
R. De Renzi,
S. Sanna, 
T. Tohyama, and
C. Ulrich
for important stimulating discussions and comments.

The main part of this work was done during my stay at the
Max Planck Institute for Solid State Research, Stuttgart,
and the Yukawa Institute, Kyoto.
I am very grateful to my colleagues
for hospitality and for stimulating atmosphere.

This work was supported 
 by the Humboldt Foundation and by
the Japan Society for Promotion of Science.


\begin{thebibliography}{99}
\bibitem{com1} La$_{2-x}$Sr$_x$CuO$_4$ is a superconductor at
$x> 0.055$, but still it is an Anderson insulator from the
point of view of the single particle physics.
\end{thebibliography}
\end{document}